\documentclass[prl,twocolumn,showpacs,preprintnumbers,amsmath,amssymb]{revtex4}
\usepackage{graphicx}% Include figure files
\usepackage{bm}% bold math
\begin{document}

\preprint{ }
\title{Asymmetry with respect to the magnetic field direction in
the interaction between the quantum states of two coupled superconducting rings}

%\thanks{}

\author{V.~I. Kuznetsov}
\email{kvi@ipmt-hpm.ac.ru}
\author{A.~A. Firsov}
\author{S.~V. Dubonos}
\author{M.~V. Chukalina}
\affiliation{Institute of Microelectronics Technology and High
Purity Materials, Russian Academy of Sciences, 142432
Chernogolovka, Moscow Region, Russia}

\date{\today}% It is always \today, today,

\begin{abstract}
The interaction between the quantum states of two aluminum
superconducting rings forming an 8-shape circular-asymmetric
microstructure was examined under a threading magnetic flux and
bias by an alternating current without a dc component. Quantum
oscillations of the rectified dc voltage $V_{dc}(B)$ as a function
of magnetic field were measured in the 8-shape microstructure at
various bias ac currents and temperatures close to critical.
Fourier and wavelet analyses of $V_{dc}(B)$ functions revealed the
presence of various combination frequencies in addition to two
ring fundamental frequencies, which suggests the interaction in
the structure. Deviation of the $V_{dc}(B)$ function from oddness
with respect to the magnetic field direction was found for the
first time.
\end{abstract}

\maketitle

The aim of the work is to study the expected interaction between
two directly coupled asymmetric rings, using a recently discovered
effect of ac voltage rectification in an asymmetric ring
\cite{q1}. Time-averaged nonzero dc voltage $V_{dc}(B)$ was
experimentally observed when a bias sinusoidal current (without a
dc component) of an amplitude close critical and frequencies up to
1 MHz was transmitted through a single circularly asymmetric
superconducting ring in a perpendicular magnetic field. The
$V_{dc}(B)$ voltage as the function of $B$ oscillates with the
period $\Delta B=\Phi_{0}/S$, where $\Phi_{0}$ is the
superconducting magnetic flux quantum and $S$ is the effective
ring area \cite{q1}. Unlike $R(B)$ oscillations in the
Little-Parks effect \cite{q2}, the $V_{dc}(B)$ function is
sign-alternating and odd with respect to the magnetic field
direction \cite{q1}. Earlier, the behavior of superconducting
loops in magnetic field was studied using $R(B)$ and $T_{c}(B)$
functions \cite{q2,q3}.

\begin{figure}
\includegraphics[width=1.0\columnwidth]{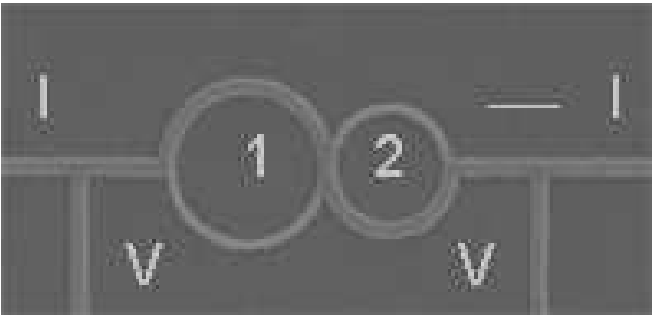}
\caption{\label{image} Scanning electron micrograph of the 8-shape
asymmetric structure with current and potential contacts. The
longitudinal mark: 2 $\mu$m.}
\end{figure}

The circular asymmetry of the structure under study (Fig.
\ref{image}) makes it possible to use the $V_{dc}(B)$ function for
the analysis of quantum behavior of the superconducting system.
The ring quantum state is fully determined by the magnitude and
direction of circulating current $I_{p}(B)$. Because the rectified
voltage $V_{dc}(B)$ in a single asymmetric ring is probably
proportional to the ring circulating current $I_{p}(B)$ \cite{q1},
measurements of the $V_{dc}(B)$ voltage can help determine the
quantum state of the ring. Measuring $V_{dc}(B)$ in the system of
different coupled asymmetric rings can be expected to provide
information on the quantum state of each ring and the interaction
between them.

The interest to the interaction in the system of superconducting
asymmetric rings is due to that a similar ring with 10 nm thick
walls at  $T<0.5T_{c}$ biased by a microwave current can
potentially be used as an element for a novel superconducting flux
qubit with quantum phase-slip centers (QPSC) \cite{q4}. Two
directly coupled rings forming an 8-shape structure can be a
prototype of two directly coupled flux qubits. In such a qubit,
quantum phase-slip centers play the role of tunnel contacts, which
are present in all known working superconducting flux qubits
\cite{q5}. Such a qubit was theoretically considered in \cite{q6},
but was not studied experimentally.

An aluminum structure $d=72$ nm thick was fabricated by Al thermal
deposition onto a silicon substrate, using lift-off process of
electron-beam lithography. The structure is 8-shape and asymmetric
structure with the wide, and narrow wires $w_{w}=0.47$ and
$w_{n}=0.24$ $\mu$m wide, respectively (Fig. \ref{image}).

Average geometrical areas of the larger and the smaller rings are
$S^{g}_{L}=14.5$ $\mu\rm m^{2}$ and $S^{g}_{S}=8.5$ $\mu\rm
m^{2}$. The parameters of the structure are the following:
$R_{4.2}=8.39$ $\Omega$ (normal-state resistance at $T=4.2$ K),
$R_{300}/R_{4.2}=2.22$ (ratio of room to helium temperature
resistances), $R_{s}=0.33$ $\Omega$ (sheet resistance), $l=25$ nm
(an effective mean free path of electrons), $T_{c}=1.324$ K
(superconducting critical temperature), $\xi(0)=170$ nm
(superconducting coherence length at $T=0$, and
$\lambda_{d}(0)=246$ nm (magnetic field penetration depth at
$T=0$).
\begin{figure}
\includegraphics[width=1.0\columnwidth]{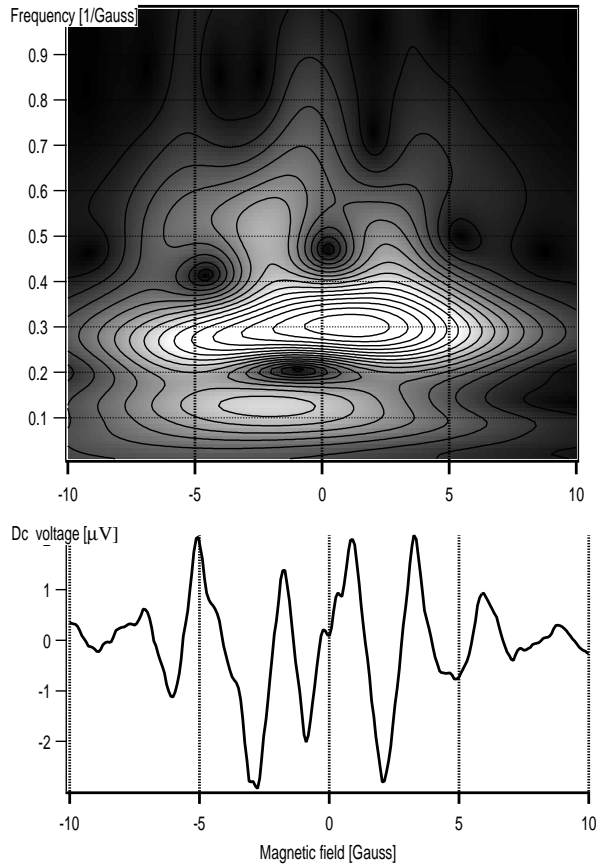}
\caption{\label{wlv} Bottom panel: The rectified voltage
$V_{dc}(B)$ in the structure at the bias sinusoidal current
$\nu=1.5$ kHz and amplitude $I_{\nu}=2.8$ $\mu$A at $T=1.306$ K;
the critical current in the zero magnetic field $I_{c} (T,B =
0)=2.27$ $\mu$A and the critical temperature $T_{c}=1.324$ K. Top
panel: The wavelet transform of the $V_{dc}(B)$ curve in the range
-40 to +40 Gauss; the wavelet spectrum amplitude as the function
of magnetic field (abscissa axis, Gauss) and frequency (ordinate
axis, ${\rm Gauss}^{-1}$) is presented along the $z$ axis as
different shades of gray (lighter areas correspond to larger
amplitudes).}
\end{figure}

In this work, we present an experimental observation of the
deviation of the $V_{dc}(B)$ function from oddness in the 8-shape
structure (Fig. \ref{image}). Figure \ref{wlv} (bottom) shows the
rectified dc voltage $V_{dc}(B)$ in the structure biased by a
sinusoidal current (without a dc component) of $\nu=1.5$ kHz and
amplitude $I_{\nu}$ close to critical. The deviation from the
$V_{dc}(B)$ function oddness is clearly seen. The fast Fourier
transform (FFT) of this not strictly odd function taken in the
whole interval of magnetic fields gives an integral picture. In
order to analyze the behavior of the $V_{dc}(B)$ function in
negative and positive magnetic fields separately, we made a
wavelet analysis of the curve in Fig. \ref{wlv} (top).

The wavelet analysis \cite{q7} allows the decomposition of a
signal into locally confined waves, wavelets. The result of the
wavelet transformation of a $f(t)$ function is determined as
$(Tf)(a,b)=\left| a\right |^{-\frac{1}{2}}\int
dtf(t)\psi(\frac{t-b}{a})$, where $\psi(t)$ is the "mother"
wavelet function. In our case, Morlet  \cite{q7} "mother" function
$\psi(t)=(\exp(i\gamma
t)-\exp(-\frac{\gamma^{2}}{2}))\exp(-\frac{t^{2}}{2})$ was used.
The parameter $\gamma$ determines an optimum number of
oscillations analyzed. The result of wavelet transformation is a
function of two parameters. The value $(\frac{1}{a})$ corresponds
to the frequency in the Fourier transformation. Each function
$\psi(\frac{t-b}{a})$ localized around $t=b$ values. Fig.
\ref{wlv} (top) gives a three-dimensional view of the $V_{dc}(B)$
curve wavelet transform.

The amplitude of the wavelet transform is shown along the $z$ axis
as different shades of gray versus magnetic field ($x$ axis) and
frequency. i.e. the reciprocal of a certain period of oscillations
($y$ axis). Lighter colors correspond to larger wavelet
amplitudes. Closed curves are lines with similar amplitude values.
The asymmetry of the wavelet amplitude with to magnetic field is
evident.

\begin{figure}
\includegraphics[width=1.0\columnwidth]{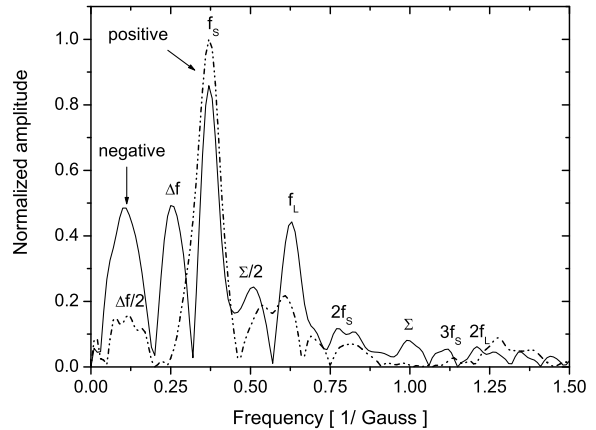}
\caption{\label{fft} The solid line presents the Fourier spectrum
(FFT) from the left part of the $V_{dc}(B)$ curve taken in the
range of negative magnetic fields (-40 to 0 Gauss) and extended to
positive magnetic field in an odd way. The dash-dotted line
presents the Fourier spectrum from the right part of the
$V_{dc}(B)$ curve taken in the range of positive magnetic fields
(0 to + 40 Gauss) and extended to negative field in an odd way.}
\end{figure}

Separate Fourier analysis of both parts of the $V_{dc}(B)$
function corresponding to negative and positive magnetic fields
was also made. To improve the resolution of the FFT spectrum, the
left (as well as the right) part of $V_{dc}(B)$ was extended to
the corresponding region of positive (negative) fields in an odd
way with respect to $B=0$. The Fourier transforms of the resulting
curves are shown in Fig. \ref{fft}. We used $2^{12}$ uniformly
distributed points in the -40 to +40 Gauss range. A set of
amplitude peaks can be seen at certain frequencies, which are
reciprocal to different oscillation periods in $V_{dc}(B)$. So,
the fundamental frequencies $f_{S}$ and $f_{L}$ corresponding to
the effective areas of the smaller and larger rings $S_{S}$ and
$S_{L}$, respectively, will be reciprocal to the corresponding
oscillation periods for the smaller and larger rings, i.e.
$f_{S}=1/ \Delta B_{S}=S_{S}/ \Phi_{0}$ and $f_{L}=1/ \Delta
B_{L}=S_{L}/ \Phi_{0}$. The values of the fundamental frequencies
expected from the structure geometry are $f^{g}_{S}=0.41$ ${\rm
Gauss}^{-1}$, $f^{g}_{L}=0.69$ ${\rm Gauss}^{-1}$. Indeed, the
Fourier spectrum exhibits peaks in the regions of these values.

In addition to the fundamental frequencies $f_{S}$ and $f_{L}$,
the spectrum displays higher harmonics of the fundamental
frequencies $f_{Sm}=mf_{S}$ and  $f_{Lm}=mf_{L}$, difference and
half-difference frequencies $f_{\Delta}=\Delta f=f_{L}- f_{S}$,
$f_{\Delta} /2=\Delta f /2$, and summation and half-summation
frequencies $f_{\Sigma}=f_{L}+f_{S}$, $f_{\Sigma}/2$.

Low-frequency peaks determined by the oscillation attenuation
field also observed. Moreover, spectral peaks are observed at
frequencies, which are combinations of the above frequencies and
higher harmonics of the fundamental frequencies. The presence of
the combination frequencies suggests interactions in the
structure.

The wavelet and FFT analyses (Figs. \ref{wlv}, \ref{fft}) show
that the response of different parts of the structure to the
magnetic field and the interaction in the structure crucially
depend on the magnetic field direction. For example, the
difference and half-difference frequencies are mainly observed in
the negative fields. When the amplitude of the bias ac current was
much larger than the magnitude of the zero-field critical current,
the magnetic asymmetry considerably decreased. These observations
cannot be explained by the presence of extra uncontrolled direct
or alternating currents and extra disregarded field. Magnetic
asymmetry could have been explained by some sort of  "freezing" of
closed currents in the structure, but this is hardly the case
because of the structure geometry. The understanding of this
strange asymmetry requires further investigations.

In conclusion, the wavelet and FFT analyses of the $V_{dc}(B)$
function in the 8-shape structure have provided information on the
quantum state of the system consisting of two rings and the
interaction between the rings. An asymmetry in the interactions in
the asymmetric superconducting structure has been observed, which
depends on the external parameters and the structure geometry.

The authors are grateful to V.~L. Gurtovoi, A.~V. Nikulov, and
V.~A. Tulin for useful discussions. The work was financially
supported in the framework of the program "Computations based on
novel physical quantum algorithms", Information Technologies and
Computer Systems Department of the Russian Academy of Sciences and
the program "Quantum Macrophysics", Presidium of the Russian
Academy of Sciences.

\end{document}